\documentclass[twocolumn,aps,prb,widetext,showpacs,superscriptaddress]{revtex4}
\usepackage{mathrsfs,bm,multirow,amsmath,amsfonts,amssymb,array,booktabs,float}
\usepackage{graphicx,times,graphics,color,epsfig}
\begin{document}
\title{
Nonequilibrium spin injection in monolayer black phosphorus}

\author{Mingyan Chen}
\affiliation{Department of Physics, Shanghai Normal University, 100 Guilin Road, Shanghai 200232, China}
\affiliation{Department of Physics and the Center of Theoretical and Computational Physics, The University of Hong Kong, Pokfulam Road, Hong Kong SAR, China}

\author{Zhizhou Yu}
\affiliation{Department of Physics and the Center of Theoretical and Computational Physics, The University of Hong Kong, Pokfulam Road, Hong Kong SAR, China}
\affiliation{The University of Hong Kong Shenzhen Institute of Research and Innovation, Shenzhen, Guangdong 518048, China.}

\author{Yin Wang}
\email{yinwang@hku.hk}
\affiliation{Department of Physics and the Center of Theoretical and Computational Physics, The University of Hong Kong, Pokfulam Road, Hong Kong SAR, China}
\affiliation{The University of Hong Kong Shenzhen Institute of Research and Innovation, Shenzhen, Guangdong 518048, China.}

\author{Yiqun Xie}
\email{yqxie@shnu.edu.cn}
\affiliation{Department of Physics, Shanghai Normal University, 100 Guilin Road, Shanghai 200232, China}
\affiliation{Center for the Physics of Materials and Department of Physics, McGill University, Montreal, PQ H3A 2T8, Canada.}

\author{Jian Wang}
\affiliation{Department of Physics and the Center of Theoretical and Computational Physics, The University of Hong Kong, Pokfulam Road, Hong Kong SAR, China}
\affiliation{The University of Hong Kong Shenzhen Institute of Research and Innovation, Shenzhen, Guangdong 518048, China.}

\author{Hong Guo}
\affiliation{Center for the Physics of Materials and Department of Physics, McGill University, Montreal, PQ H3A 2T8, Canada.}
\affiliation{Department of Physics and the Center of Theoretical and Computational Physics, The University of Hong Kong, Pokfulam Road, Hong Kong SAR, China}

\begin{abstract}
Monolayer black phosphorus (MBP) is an interesting emerging electronic material with a direct band gap and relatively high carrier mobility. In this work we report a theoretical investigation of nonequilibrium spin injection and spin-polarized quantum transport in MBP from ferromagnetic Ni contacts, in two-dimensional magnetic tunneling structures. We investigate physical properties of the spin injection efficiency, the tunnel magnetoresistance ratio, spin-polarized currents, charge currents and transmission coefficients as a function of external bias voltage, for two different device contact structures where MBP is contacted by Ni(111) and by Ni(100). While both structures are predicted to give respectable spin-polarized quantum transport, the Ni(100)/MBP/Ni(100) trilayer has the superior property where the spin injection and magnetoresistance ratio maintains almost a constant value against the bias voltage. The nonequilibrium quantum transport phenomenon is understood by analyzing the transmission spectrum at nonequilibrium.
\end{abstract}
\maketitle

Two dimensional (2D) materials have received extensive investigations in recent years for possible applications in logic devices, photonic systems, solar cells, transparent substrates and perhaps most interestingly, flexible and wearable consumer electronics.\cite{nnano-review} The thin layer of 2D material makes it a natural choice for producing flexible structures due to their out of plane flexibility. Many 2D materials have strong covalent bonds and diverse electronic structures - properties which are needed for reliable and durable applications.

So far, several 2D materials have been fabricated successfully including the celebrated graphene,\cite{graphene1i, graphene2i, graphene3i, graphene4i} various 2D transition-metal dichalcogenides,\cite{TMDC1i, TMDC2i, TMDC3i} and the monolayer black phosphorus (MBP).\cite{Li, Hong, Liu, jiwei, Rodin} In particular, as one of the newest members of 2D material family, MBP is very interesting in several aspects. First, different from transition-metal dichalcogenides, black phosphorus is made of a single atomic specie, phosphorus. Second, different from graphene, the phosphorus atoms in MBP are not all located in a plane but form a buckled hexagonal structure by covalence bonds and few-layer black phosphorus has an ideal direct band-gap, a property that is very important for optoelectronics. Third, MBP has an intrinsic band gap and graphene does not. Though lower than that of graphene, few-layer black phosphorus has respectable mobilities of $\sim 1000$cm$^2$V$^{-1}$s$^{-1}$ as reported experimentally.\cite{Li}

While the materials properties make MBP very interesting and potentially important for emerging flexible electronics, another critical issue is to achieve low power operation. In this regard, one notes that the energy scale of spin dynamics is typically many orders of magnitude smaller than that of charge dynamics, and low power electronics operation can thus be achieved in spintronics devices whose operation principle is based on spin dynamics.\cite{Wolf, ASLD} Existing and well studied spintronic systems include magnetic random access memory, all spin logic device, and magnetic sensors. The tunnel magnetoresistance (TMR) is one of the most important spintronics phenomena observed in magnetic tunnel junctions (MTJ) which are made of two ferromagnetic contacts sandwiching a nanometer thin insulating material. The tunneling current is large when magnetic moments of the two magnetic contacts are in parallel configuration (PC) and it is small when they are in antiparallel configuration (APC). An important device merit is the TMR ratio and much theoretical and experimental efforts have been devoted to create MTJs with different ferromagnetic metals and insulating materials in order to generate a large ratio. While materials such as MgO and Al$_2$O$_3$ are the most popular barrier materials in practical MTJs,\cite{Zhang, Waldron, Parkin, Yuasa, Jeon} 2D materials graphene\cite{graphene1,graphene2} and transition-metal dichalcogenides\cite{TMDC1,TMDC2} have also been investigated in this context.

Given the huge interests in 2D nano-materials and the lack of knowledge about spin injection in MBP, in this work we investigate 2D MTJs consisting of a MBP as the tunnel barrier sandwiched by Ni contacts\cite{Ni} based on a state-of-the-art theoretical approach where density functional theory (DFT)\cite{DFT} is combined with the Keldysh nonequilibrium Green's function (NEGF) theory.\cite{NEGF}  We are interested in understanding the nonequilibrium spin injection property of MBP driven by a finite external bias voltage. It was known that in the operational bias range the TMR ratio monotonically diminishes to zero for MgO based MTJs.\cite{Waldron} For MBP, we found that the spin injection and TMR ratio maintains a relatively large value and independent of a significant range of bias. We investigate physical properties of the spin injection efficiency, the tunnel magnetoresistance ratio, spin-polarized currents, charge currents and transmission coefficients as a function of external bias voltage, for two different device contact structures where MBP is contacted by Ni(111) and by Ni(100). Both structures are predicted to give respectable spin-polarized quantum transport, the Ni(100)/MBP/Ni(100) trilayer has the superior property where the spin injection and magnetoresistance ratio maintains a large and relatively constant value against the bias voltage. The nonequilibrium quantum transport phenomenon is understood by analyzing the transmission spectrum.

\begin{figure}[tbp]
\includegraphics[width=\columnwidth]{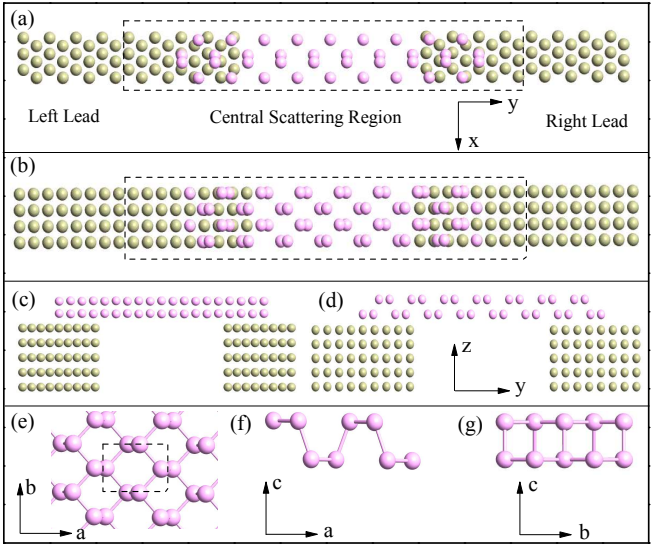}
\caption{The top view of the atomic structure of: (a) the Ni(111)/MBP/Ni(111) and (b) the Ni(100)/MBP/Ni(100) MTJ. (c) and (d) are the side views of the central scattering region of Ni(111) and Ni(100) MTJs. The 2D structures periodically extend in the $x$ direction, the current flows in the $y$ direction. (e) The top view, (f) and (g) the side view of the MBP, $a$ and $b$ directions correspond to the long and short direction of the MBP in real space, respectively. Yellow spheres denote Ni atoms and pink spheres denote P atoms.}\label{fig1}
\end{figure}

Fig.~\ref{fig1} plots the two atomic models of the 2D Ni/MBP/Ni MTJ which we investigate, one contacted by the Ni(111) surface and the other by the Ni(100) surface. Because MBP is a 2D material, structures of Fig.~\ref{fig1}(c,d) periodically extends in the $x$ direction (current flows in $y$ direction) with a periodicity of 4.316~\AA~for the MTJ with Ni(111); and 3.524~\AA~for the MTJ with Ni(100). The lattice constant of MBP along the $a$ (long) and $b$ (short) directions are 4.58~\AA~and 3.32~\AA, respectively [see Fig.~\ref{fig1}(e,f,g)].\cite{jiwei} To build a periodic structure along the $x$ direction for the Ni/MBP/Ni MTJ, the MBP is homogeneously strained by about $\pm 6\%$ to match the Ni lattice. Because we are interested in 2D device structures, the magnetic electrodes are made of Ni slabs consists of five layers of Ni atoms, and the electrodes extend to $y=\pm \infty$ where bias voltages are applied and electric current collected. In our two-probe MTJ model, the MBP material overlaps with the Ni slab surface on either end to form a current-in-plane configuration [see Fig.~\ref{fig1}(c,d)],\cite{gong} which is similar to a device structure in a recent experiment.\cite{kamalakar} The distance between the MBP and the Ni slab surface is obtained by DFT total energy relaxation\cite{DFT} which produced an optimized value from the bottom sub-layer of the MBP to the Ni(111) and Ni(100) surfaces to be 2.0~\AA~and 1.95~\AA, respectively. The distance between the two Ni electrodes in the $y$ direction, namely the length of the MBP not overlapping with the Ni [see Fig.~\ref{fig1}(c,d)], are 18.2~\AA~for the Ni(111), and 21.46~\AA~for the Ni(100). Finally, in the numerical calculations a vacuum region of 20~\AA~in the $z$ direction is included in the 2D MTJ supercell to isolate any possible spurious interaction between periodical images of the supercell. In the relaxation, DFT as implemented in the VASP package\cite{VASP} was adopted and the exchange-correlation energy was treated by the projector augmented wave of the Perdew-Burke-Ernzerhof\cite{PAW} with an energy cutoff of 500 eV. The Brillouin zone was sampled with a 10$\times$8$\times$1 mesh of the Monkhorst-Pack k-points.\cite{Kpoint}

Having determined the atomic structures of the 2D MTJ, nonequilibrium spin-polarized quantum transport properties were calculated by the NEGF-DFT quantum transport package Nanodcal.\cite{NEGFDFT} The essential ingredients of the NEGF-DFT formalism are consisted of: (i) For a given density matrix, calculating the Hamiltonian of the two-probe open device by a DFT-like self-consistent field theory where the external voltages provide electrostatic boundary conditions when solving the Hartree potential; (ii) For a given Hamiltonian, calculating the density matrix by NEGF; (iii) The procedure is repeated until a self-consistent solution of both the Hamiltonian and the NEGF are obtained. Afterward, quantum transport properties are calculated by the final converged NEGF, including the transmission spectra at finite bias voltage $V$, $T(E,V)$ where $E$ is the electron energy; and transport current which is obtained by integrating over the bias window $-V/2 \leq E\leq +V/2$, \emph{i.e.} $I \sim \int_{-V/2}^{+V/2} T(E,V) dE$. Clearly, due to spin polarization the quantities $T(E,V)$ and $I$ all possess spin quantum index.
We refer interested readers to the original literature Ref.~\onlinecite{NEGFDFT} for further technical details of NEGF-DFT. In our calculations, double-zeta polarized atomic orbital basis was used to expand the physical quantities;\cite{DZP} the exchange-correlation were treated at the level of local spin density approximation;\cite{LDA,LDA2,foot1} atomic cores are defined by the standard norm conserving nonlocal pseudopotentials;\cite{Troullier} and $300\times1\times1$ k-points were used to calculate the electric current. We have calculated magnetic moments of the Ni electrodes and the obtained values for the first three layers in the unit of $\mu_B$ are 0.632, 0.661, and 0.629 for (111) surface, and 0.743, 0.626, and 0.637 for (100) surface, which are in excellent agreement with those reported in the literature\cite{MM,MM2} with difference less than 2.5\% compared to the values in Ref.~\onlinecite{MM2}.

For MTJs with the Ni(111) surface, the current flows along the $b$ direction in the pure MBP region; for MTJs with Ni(100), the current flows along the $a$ direction [see Fig.~\ref{fig1}(e,f,g)]. In the following we analyze two important device merits, the TMR ratio defined as $TMR \equiv (I_{PC}-I_{APC})/I_{APC}$; and the spin-injection efficiency (SIE) defined as
$\eta\equiv \frac{|I_{\uparrow}-I_{\downarrow}|}{|I_{\uparrow}+I_{\downarrow}|}$. Here, $I_{PC}, I_{APC}$ are the charge currents for situations where the magnetic moments of the two Ni contacts are in PC or APC, respectively; $I_{\uparrow}, I_{\downarrow}$ denote the spin-polarized current contributed by the spin-up and -down channels respectively, and total charge current is $I_{\uparrow}+I_{\downarrow}$. At zero bias when all currents vanish, we use transmission coefficient at the Fermi level to calculate TMR and $\eta$. Physically, the TMR ratio measures the sensitivity of the MTJ device with respect to the magnetic configuration, and SIE measures the extent of spin polarization in the transport current.

\begin{figure}[tbp]
\includegraphics[width=\columnwidth]{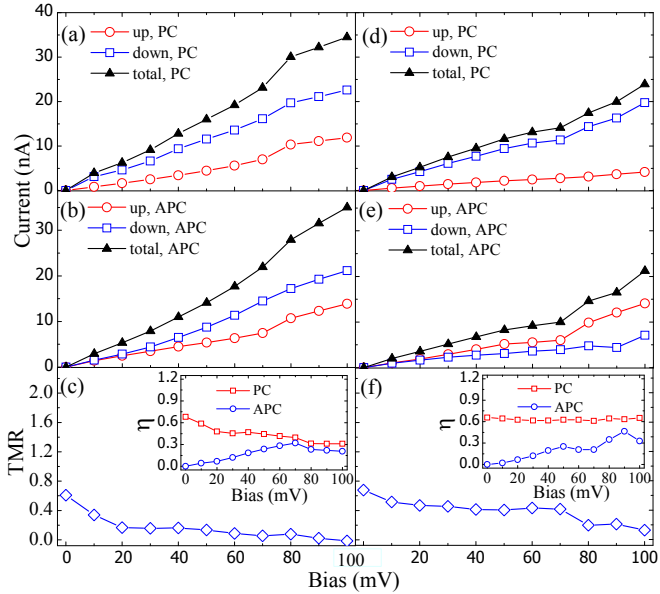}
\caption{Panels (a,b,c) are for Ni(111) contacted MTJ. (a) I-V curves for PC and (b) I-V curves for APC; (c) TMR and SIE (inset) versus bias. Panels (d,e,f) are for Ni(100) contacted MTJ. (d) I-V curves for PC and (e)I-V curves for APC; (f) TMR and SIE (inset) versus bias. Note that the Ni(100)/MBP/Ni(100) MTJ has a significantly higher TMR ratio which is essentially constant versus bias up to about 70~mV.}\label{fig2}
\end{figure}

Fig.~\ref{fig2}(a,b) and Fig.~\ref{fig2}(d,e) present the calculated spin-polarized currents and total currents of the MTJs with Ni(111) and with Ni(100) respectively, versus the bias voltage up to 100~mV. For both Ni(111) and Ni(100) MTJs, the total current $I_{PC, APC}$ (black curves with up-triangles) essentially increases linearly with bias up to about 70~mV at which  nonlinearity appears as indicated by a more rapid change of $I_{PC, APC}$. As for the spin-polarized currents (curves with blue squares and red circles), we found $I_{\downarrow}> I_{\uparrow}$ for both PC and APC in the Ni(111) system [Figs.\ref{fig2}(a,b)]. On the other hand, for Ni(100) systems $I_{\downarrow}> I_{\uparrow}$ for PC [Fig.~\ref{fig2}(d)] while $I_{\downarrow} <  I_{\uparrow}$ for APC [Fig.~\ref{fig2}(e)]. From spin-polarized currents one obtains the SIE coefficient $\eta$ which is presented in the inset of Fig.~\ref{fig2}(c,f). A distinct feature of $\eta$ is observed for the PC case of the Ni(100) MTJ [red squares in the inset of Fig.~\ref{fig2}(f)], namely it essentially maintains a constant SIE value of 60\% independent of bias up to 100~mV. Using the calculated total current for PC and APC, we obtain the TMR ratio for the two MTJs as shown in Fig.~\ref{fig2}(c,f). At the zero bias limit, TMR is 61\% and 67\% for Ni(111) and Ni(100) devices, respectively. Starting from these values, the bias voltage suppresses TMR gradually and eventually to zero at about 100~mV. A most interesting result is found for the Ni(100) MTJ: it maintains a stable TMR $\approx 40\%$ up to 70~mV bias. Being able to maintain a substantial and stable TMR versus bias is very important for practical applications: it allows one to tune charge currents with bias while maintaining the same TMR ratio. Overall, our numerical results thus suggest that the 2D Ni(100)/MBP/Ni(100) MTJ is a better system by the two device merits: it has larger TMR for the full bias range and maintains a stable TMR up to 70~mV; it also has a stable and higher SIE against the external bias.

\begin{figure}[tbp]
\includegraphics[width=\columnwidth]{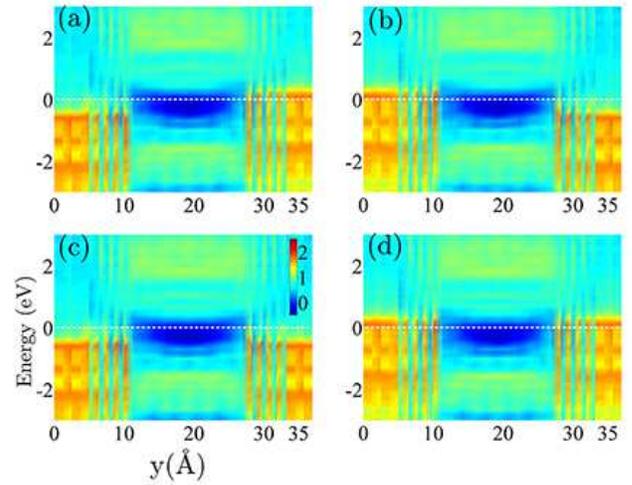}
\caption{Project density of states (PDOS) by different colours in logarithmic scale along the transport direction (y direction) of the Ni(111)/MBP/Ni(111) MTJ at equilibrium. (a) Spin up states in APC, (b) spin down states in APC, (c) spin up states in PC, and (d) spin down states in PC. All the sub-figures have the same axes as (c), color coding values are given by the vertical bar in (c). White dashed lines indicate the Fermi level.}\label{fig3}
\end{figure}

\begin{figure}[tbp]
\includegraphics[width=\columnwidth]{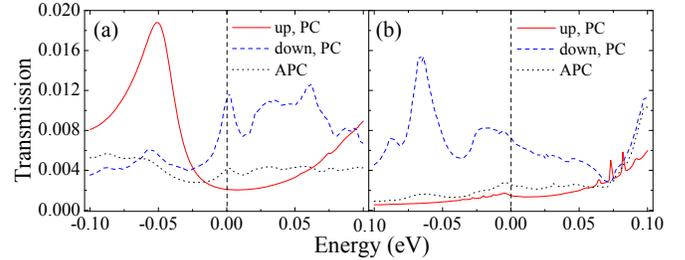}
\caption{Zero bias transmission coefficient versus electron energy in PC and APC of: (a) the Ni(111)/MBP/Ni(111) junction and (b) the Ni(100)/MBP/Ni(100) junction. The Fermi level is at the energy zero.}\label{fig4}
\end{figure}

Having presented the calculated numerical results, we now provide an more intuitive understanding of the quantum transport through Ni/MBP/Ni junction from the project density of states (PDOS) of the MTJ plotted along the transport direction.\cite{PDOS} Figs.~\ref{fig3}(a-d) plot the PDOS of the Ni(111)/MBP/Ni(111) junction by different colours in logarithmic scale, the PDOS of the Ni(100)/MBP/Ni(100) junction can be analyzed similarly. Several observations are in order. (i) The calculated Fermi levels go through the band gap of MBP (dark blue region, from 10~\AA~to nearly 30~\AA~in the figure),  indicating the tunneling transport mechanism. (ii) The Fermi levels locate at about 300 meV below the conduction band bottom of MBP, hence the MTJ works by direct tunneling as long as the bias voltage is less than this value. Our nonquilibrium transport calculations are performed below 100 mV. (iii) In APC [Fig.\ref{fig3} (a,b)], the spin-up (spin-down) electrons flow from the left Ni electrode into the MBP with a smaller (larger) density of states, and go out of the MBP with a larger (smaller) density of state via the right Ni electrode. The smaller density of states at one of the Ni electrodes limit transport, provides the MTJ with a high resistance state. (iv) In PC, the spin-up electrons have smaller density of states at both electrodes [Fig.\ref{fig3} (c)] while the spin-down electrons have larger density of states to dominate transport [Fig.\ref{fig3} (d)]. In general, from the PDOS in Fig.~\ref{fig3} one can intuitively understand how an electron traverses via the Ni electrodes by tunneling through the MBP barrier.

Next, we provide an understanding of the TMR and SIE by analyzing transmission spectra. Figs.\ref{fig4}(a,b) present the calculated transmission coefficients versus electron energy at zero bias for the Ni(111) and Ni(100) MTJs, respectively. For both structures, due to the geometric mirror symmetry of the atomic structure respect to the middle plane of the scattering region, the calculated transmission coefficients in APC is exactly the same for spin-up and -down channels, thus there is only one APC curve (black dotted line) in Figs.\ref{fig4}(a,b). This also serves as a very strict verification of the numerical accuracy in our calculations. For PC, there are two transmission curves for the two spin channels. Clearly, at the Fermi level the spin-down channel gives significantly larger contribution (blue dashed line) than the spin-up channel to the total transmission for both MTJs, and this gives rise to the relatively large values of TMR at zero bias [see Fig.~\ref{fig2}(c,f)]. For SIE, the same mechanism gives rise to its relatively large value at zero bias in PC; but the transmission symmetry in APC produces a zero SIE when there is no bias. The calculated transmission coefficients qualitatively agree with the PDOS analysis above: in PC the spin down electrons have larger density of states at both Ni leads, therefore the transmission of spin down channels in PC is larger; in APC symmetrical density of states are observed for spin down and spin up states, therefore equal transmission coefficients can be expected.

\begin{figure}[tbp]
\includegraphics[width=\columnwidth]{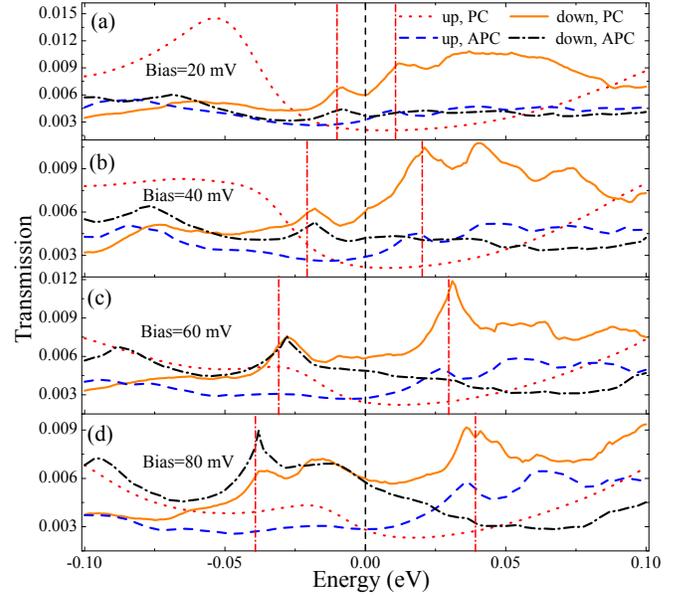}
\caption{Transmission coefficient versus electron energy for the  Ni(111)/MBP/Ni(111) MTJ at different bias voltages $V$: (a) $V=20$mV; (b) $40$mV; (c) $60$mV; (d) $80$mV. The bias window in each panel is between the two red vertical dashed-dotted lines, and the zero energy point is set at the middle of the bias window.}\label{fig5}
\end{figure}

\begin{figure}[tbp]
\includegraphics[width=\columnwidth]{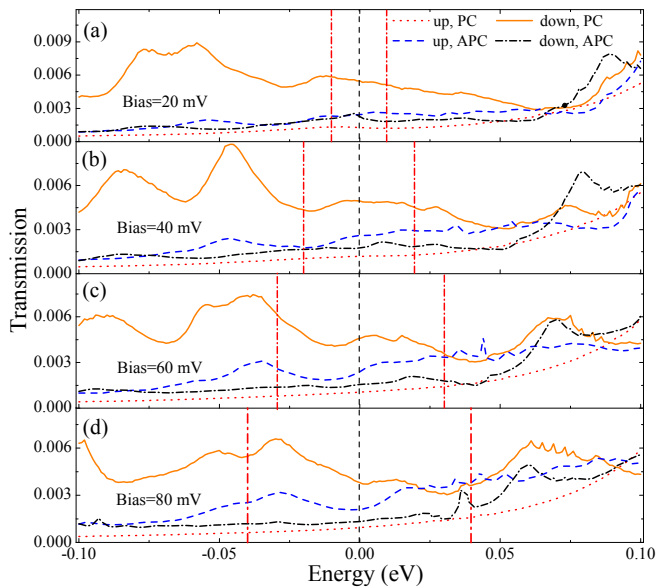}
\caption{Transmission coefficient versus electron energy for the  Ni(100)/MBP/Ni(100) MTJ at different bias voltages $V$: (a) $V=20$mV; (b) $40$mV; (c) $60$mV; (d) $80$mV. The bias window in each panel is between the two red vertical dashed-dotted lines, and the zero energy point is set at the middle of the bias window.}\label{fig6}
\end{figure}

Having analyzed the equilibrium transport properties from PDOS and transmission, we now turn to nonequilibrium quantum transport phenomenon by analyzing transmission spectra $T(E,V)$ at a finite bias $V$ from which the current is obtained by integrating over the bias window $-V/2 \leq E\leq +V/2$, as discussed above. Fig.~\ref{fig5} presents $T(E,V)$ versus $E$ for the Ni(111) MTJ at four voltages from 20~mV to 80~mV. There are four curves in each panel because a finite bias breaks the geometric symmetry and the spin-up and -down transmissions in APC no longer equal to each other. A general observation is that the spin-down channel contributes a larger transmission than the spin-up channel in the bias window for both PC and APC, explaining the result of $I_{\downarrow}> I_{\uparrow}$ for the Ni(111) MTJ [Figs.\ref{fig2}(a,b)].
In addition, for PC the spin-up transmission has a broad peak located at around $E=-0.05$~eV [red dotted line in Fig.~\ref{fig5}(a)], and the spin-down transmission has a broad peak centered at about $E=0.04$~eV [brown solid line in Fig.~\ref{fig5}(a)]. These broad peaks are diminished when the bias voltage is increased. In APC, the spin-down transmission increases substantially with the bias since a peak below the Fermi energy shifts to enter the bias window to contribute. Therefore, an increasing APC transmission and a decreasing PC transmission give rise to the reduction of TMR versus $V$, explaining the monotonic decreasing curve for the Ni(111) MTJ [see Fig.~\ref{fig2}(c)].

We also carried out the same analysis for the Ni(100) MTJ by calculating the transmission spectra $T(E,V)$, shown in Fig.~\ref{fig6}. In PC and near the Fermi level, the finite bias has only a weak influence on both spin-down and -up channels. For APC the spin-up transmission is greater than the spin-down transmission, hence the APC current is dominated by the spin-up channel [see Fig.~\ref{fig2}(e)]. Moreover, as the voltage is increased, a peak of the spin-up channel (blue dashed line) first appears in the range of $-0.05$ to $-0.02$ eV, then shifts toward the Fermi level, and finally enters the bias window at 80~mV; at the same time, a peak of the spin-down channel (black dashed-dotted line) shifts from $+0.08$ eV downward, and finally enters the the bias window at 80~mV. Hence when $V$ reaches above 70~mV, these two peaks contribute to the total current of APC, and lead to the abrupt decrease of TMR at 70~mV for the Ni(100)MTJ [see Fig.~\ref{fig2}(f)].

From these results, we see that the nonequilibrium spin injection into MBP with the current-in-plane configuration is quite significant for the two devices we investigated. Experimentally, Ref.~\onlinecite{kamalakar} recently reported measurements of transistor properties for black phosphorus contacted by ferromagnetic alloys. While the experimental measurements were not spin-polarized, the authors used a semi-classical spin diffusion model to predict that the magnetoresistance effect can be observed in their devices.\cite{kamalakar} Given the rapid progress in 2D fabrication and characterization techniques, magnetic tunnel junction devices down to a single black phosphorus layer, should be within the reach in the near future.

In summary, using a state-of-the-art first principles approach, we have investigated the properties of nonequilibrium spin injection in 2D MBP based magnetic tunnel junction. The spin injection efficiency, tunnel magnetoresistance ratio, spin-polarized currents, charge currents and transmission coefficients as a function of external bias voltage were predicted. While both structures where MBP is contacted by Ni(111) and Ni(100), are found to give respectable spin-polarized quantum transport, the Ni(100)/MBP/Ni(100) trilayer has the desired property where the spin injection and magnetoresistance ratio are not only large, namely TMR$\sim 40\%$ and SIE$\sim 60\%$ (PC), but also maintains at these values for a broad voltage range. The nonequilibrium quantum transport properties were analyzed and understood by investigating the transmission spectra at nonequilibrium. The results suggest that the Ni(100)/MBP/Ni(100) trilayer should be a promising candidate for 2D flexible spintronics system.

\section*{Acknowledgements}
Y.W. is grateful to Dr. Lei Liu for useful discussions regarding the use of the Nanodcal transport package; and to Dr. Lingling Tao for discussions on various concepts of spintronics. We thank Compute-Canada for the computation resources. This work is supported by the University Grant Council (Contract No. AoE/P-04/08, Y.W. and J.W.) of the Government of HKSAR, NSFC under Grand No. 11404273 (Y.W.) and 11374246 (J.W.), and NSERC of Canada (H.G.).

\end{document}